\documentclass{webofc}
\usepackage[varg]{txfonts}   
\DeclareUnicodeCharacter{2212}{-} 
\begin{document}
\title{Baikal-GVD: status and prospects}
%
%

\author{\firstname{A.D.} \lastname{Avrorin}\inst{1} \and
\firstname{A.V.} \lastname{Avrorin}\inst{1} \and
\firstname{V.M.} \lastname{Aynutdinov}\inst{1} \and
\firstname{R.} \lastname{Bannash}\inst{7} \and
\firstname{I.A.} \lastname{Belolaptikov}\inst{2} \and
\firstname{V.B.} \lastname{Brudanin}\inst{2} \and
\firstname{N.M.} \lastname{Budnev}\inst{3} \and
\firstname{A.A.} \lastname{Doroshenko}\inst{1} \and
\firstname{G.V.} \lastname{Domogatsky}\inst{1} \and
\firstname{R.} \lastname{Dvornický}\inst{2,8} \and
\firstname{A.N.} \lastname{Dyachok}\inst{3} \and
\firstname{Zh.-A.M.} \lastname{Dzhilkibaev}\inst{1} \and
\firstname{L.} \lastname{Fajt}\inst{2,8,9} \and
\firstname{S.V.} \lastname{Fialkovsky}\inst{5} \and
\firstname{A.R.} \lastname{Gafarov}\inst{3} \and
\firstname{K.V.} \lastname{Golubkov}\inst{1} \and
\firstname{T.I.} \lastname{Gres}\inst{3} \and
\firstname{Z.} \lastname{Honz}\inst{2} \and
\firstname{K.G.} \lastname{Kebkal}\inst{7} \and
\firstname{O.G.} \lastname{Kebkal}\inst{7} \and
\firstname{E.V.} \lastname{Khramov}\inst{2} \and
\firstname{M.M.} \lastname{Kolbin}\inst{2} \and
\firstname{K.V.} \lastname{Konischev}\inst{2} \and
\firstname{A.P.} \lastname{Korobchenko}\inst{2} \and
\firstname{A.P.} \lastname{Koshechkin}\inst{1} \and
\firstname{V.A.} \lastname{Kozhin}\inst{4} \and
\firstname{V.F.} \lastname{Kulepov}\inst{5} \and
\firstname{D.A.} \lastname{Kuleshov}\inst{1} \and
\firstname{M.B.} \lastname{Milenin}\inst{5} \and
\firstname{R.A.} \lastname{Mirgazov}\inst{3} \and
\firstname{E.R.} \lastname{Osipova}\inst{4} \and
\firstname{A.I.} \lastname{Panfilov}\inst{1} \and
\firstname{L.V.} \lastname{Pan'kov}\inst{3} \and
\firstname{D.P.} \lastname{Petukhov}\inst{1} \and
\firstname{E.N.} \lastname{Pliskovsky}\inst{2} \and
\firstname{M.I.} \lastname{Rozanov}\inst{6} \and
\firstname{E.V.} \lastname{Rjabov}\inst{3} \and
\firstname{V.D.} \lastname{Rushay}\inst{2} \and
\firstname{G.B.} \lastname{Safronov}\inst{2} \and
\firstname{F.} \lastname{Simkovic}\inst{2,8} \and
\firstname{B.A.} \lastname{Shoibonov}\inst{2} \and
\firstname{A.G.} \lastname{Solovjev}\inst{2} \and
\firstname{M.N.} \lastname{Sorokovikov}\inst{2} \and
\firstname{M.D.} \lastname{Shelepov}\inst{1} \and
\firstname{O.V.} \lastname{Suvorova}\inst{1} \and
\firstname{I.} \lastname{Shtekl}\inst{2,9} \and
\firstname{V.A.} \lastname{Tabolenko}\inst{3} \and
\firstname{B.A.} \lastname{Tarashansky}\inst{3} \and
\firstname{S.A.} \lastname{Yakovlev}\inst{7} \and
\firstname{A.V.} \lastname{Zagorodnikov}\inst{3} \and
\firstname{V.L.} \lastname{Zurbanov}\inst{3}
}

\institute{Institute for Nuclear Research, Moscow, 117312 Russia
\and
Joint Institute for Nuclear Research, Dubna, 141980 Russia
\and
Irkutsk State University, Irkutsk, 664003 Russia
\and
Institute of Nuclear Physics, Moscow State University, Moscow, 119991 Russia
\and
Nizhni Novgorod State Technical University, Nizhni Novgorod, 603950 Russia
\and
St. Petersburg State Marine Technical University, St. Petersburg, 190008 Russia
\and
EvoLogics, Berlin, Germany
\and
Comenius University, Bratislava, Slovakia
\and
Czech Technical University, Prague, Czech Republic
          }

\abstract{%
Baikal-GVD is a next generation, kilometer-scale neutrino
telescope under construction in Lake Baikal. It is designed
to detect astrophysical neutrino fluxes at energies from a
few TeV up to 100 PeV. GVD is formed by multi-megaton
subarrays (clusters). The array construction was started in 2015
by deployment of a reduced-size demonstration cluster named "Dubna"
. The first cluster in it's baseline configuration was deployed
in 2016, the second in 2017 and the third in 2018.
The full-scale GVD will be an array of $\sim$10.000 light sensors
with an instrumented volume about of 2 cubic km. The first
phase (GVD-1) is planned to be completed by 2020-2021.
It will comprise 8 clusters with 2304 light sensors in total.
We describe the design of Baikal-GVD and present selected
results obtained in 2015 – 2017.
}
\maketitle
\section{Introduction}
\label{intro}
The deep underwater neutrino telescope Baikal Gigaton Volume Detector
(Baikal-GVD) is currently under construction in Lake Baikal [1].
Baikal-GVD is formed by a three-dimensional lattice
of optical modules (OMs) arranged at vertical load-carrying
cables to form strings. The telescope has a modular structure
and consists of functionally independent clusters -
sub-arrays comprising a total of 288 OMs each and connected to
shore by individual electro-optical cables.
The first, reduced size cluster named
“Dubna” has been deployed in Lake Baikal and was operated during
2015.
In April 2016, this array has been upgraded to the baseline configuration
of a GVD-cluster, which comprises 288 optical
modules attached at 8 strings at depths from 750 m to 1275 m.
In 2017 and 2018 the second and the third GVD-clusters
were deployed, increasing the total number of
operating optical modules to 864 OMs. During Phase-1 of
Baikal-GVD implementation an array consisting of eight clusters
will be deployed by 2020-2021. Since each GVD-cluster represents a
multi-megaton scale Cherenkov detector, studies of neutrinos of
different origin are allowed with early stages of construction.

\section{Detector}
\label{sec-1}
The detector instruments the deep water of Lake Baikal
with optical modules – pressure resistant glass spheres
equipped with photomultiplier
tubes (PMT) Hamamatsu R7081-100 with photocathode diameter of
10” and  a quantum efficiency of $\sim$35\% \cite{OM}. The PMTs record the
Cherenkov radiation from secondary particles produced in
interactions of high-energy neutrinos inside or near the
instrumented volume. From the arrival times of light at the
PMTs and from the amount of light, direction and energy of the
incoming neutrinos are derived. Baikal-GVD in it's 2018 design
consists of three clusters – each of them with 288 optical modules
(see Figure 1). A cluster
comprises eight vertical strings attached to the lake floor:
seven side strings on a radius of 60 m around a central one.
Each string carries 36 OMs, arranged at depths between 735
and 1260 meters (525 m instrumented length). The vertical
spacing between the OMs along a string is 15 m. The OMs on each
string are
functionally combined in 3 sections. A section comprises
12 OMs with data processing and communication electronics and
forms a detection unit (DU) of the array. All analogue
signals from the PMTs are digitized, processed in the
sections and sent to shore if certain trigger conditions
(e.g. a minimum number of fired PMTs) are fulfiled \cite{DAQ}.

\begin{figure}[h]
\centering
\includegraphics[width=6cm,clip]{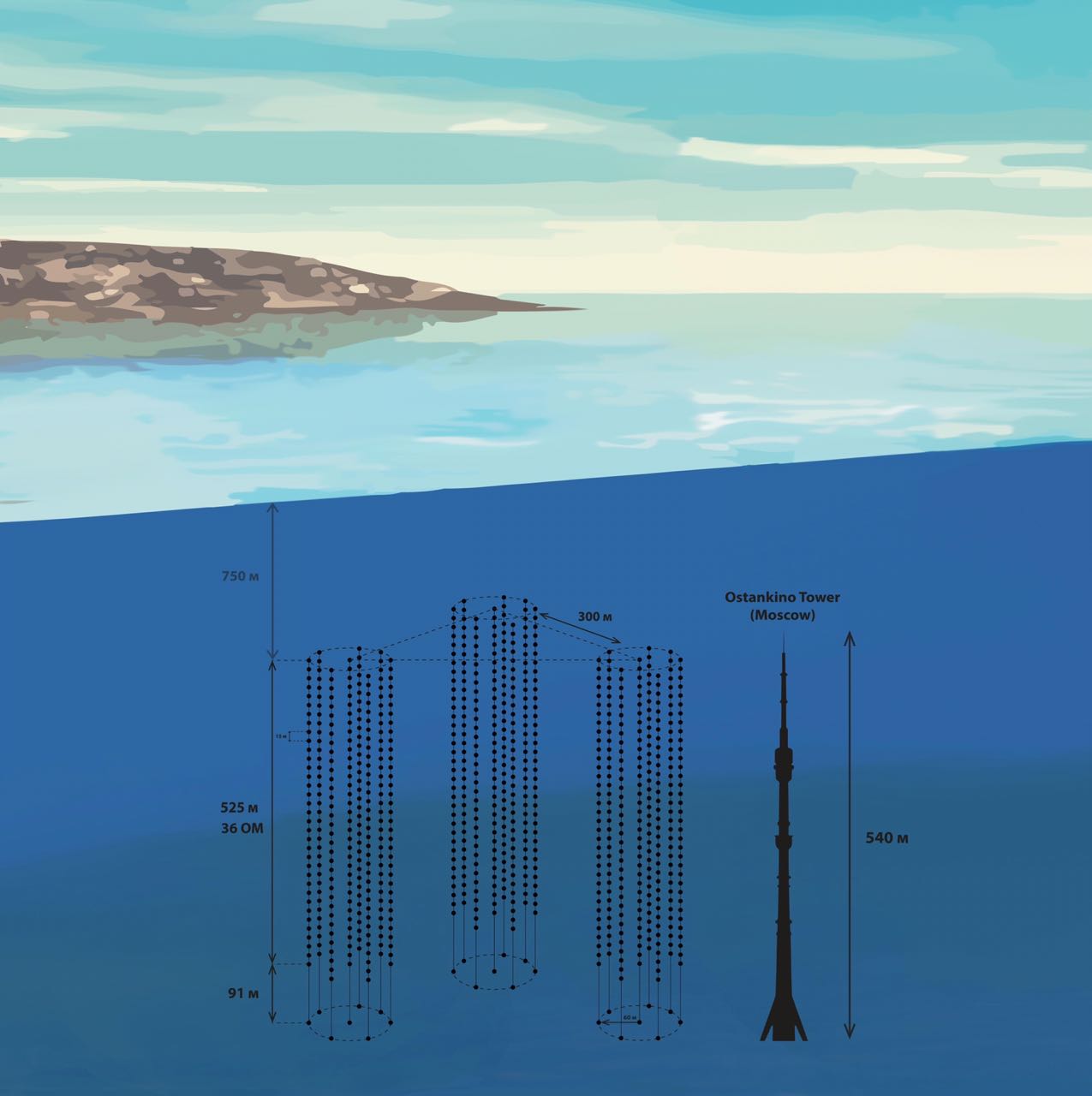}
\caption{Artist's view of GVD-2018, compared to the Moscow
television tower.}
\label{fig-1.1}       
\end{figure}

The clusters are connected to shore ($\sim$3.5 km distance) via
a network of cables for electrical power and high-bandwidth
data communication. The shore station provides power, detector
control and readout, computing resources and a high-bandwidth
internet connection to the data repositories. The overall
design allows for a flexible and cost-effective implementation
of Baikal-GVD. The large detection volume,
combined with high angular and energy resolution and moderate
background conditions in the fresh lake water allows for efficient
study of cosmic neutrinos, muons from charged cosmic rays and search for
exotic particles. It is also an attractive platform
for environmental studies. 

\section{Selected results}
\label{sec-2}
\subsection{Search for muon neutrinos}
\label{sec-2.1}
A search for upward moving muon neutrinos was performed with
the data collected
by the first cluster of the telescope in 2016. A set of 70 runs in
which the detector demonstrated stable behaviour was chosen. The total
exposition time of the analysed sample is close to 33 days.
Precise measurements of optical module positions, timing and amplitude
calibrations were available for the whole year 2016 and were applied
for the present analysis.

\begin{figure}[h]
\includegraphics[width=6.cm,height=5.3cm]{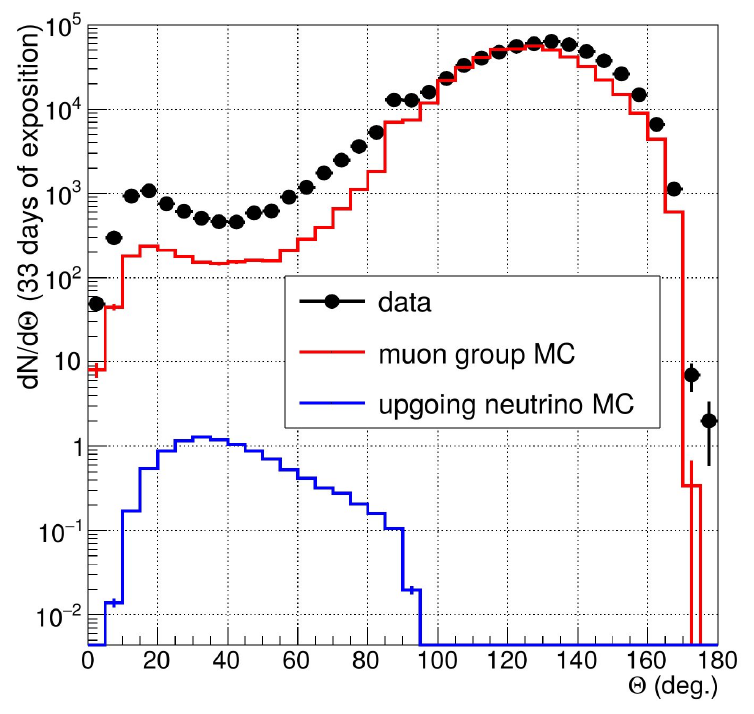}
\hfill
\includegraphics[width=6cm,clip]{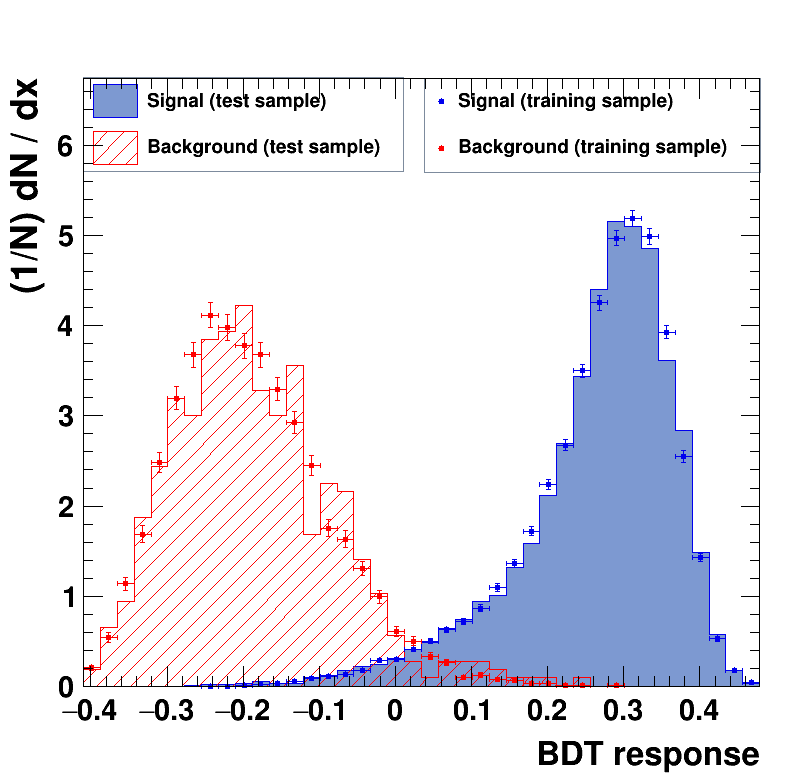}
\caption{Left: Angular distributions of atmospheric muons accumulated
over 33 days of exposition and compared to normalised Monte Carlo for
atmospheric muons. The figure also shows the angular distribution
for muons from upward moving atmospheric neutrinos.
 Right: Distribution of BDT discriminant value for signal
 (blue) and background (red) samples.}
\label{muons_33days}       
\end{figure}

Reconstruction of muons is performed in two
stages. The collection of PMT signals in each event includes noise
pulses due to PMT dark current and light background from the lake water.
Such pulses appear with 20-100 kHz per OM depending on season and depth.
These pulses are rejected with the noise suppression procedure. Pulses
are combined by the causality requirement:
$\vert t_{i}-t_{j} \vert \le {{\Delta R_{ij}} \over {c_w}} + t_{s}$,
where $t_{i}$, $t_{j}$ are pulse times, $\Delta R_{ij}$
is distance between modules,
$c_{w}$ is the speed of light in water and $t_{s}$ = 10 ns.
A simple track position
and direction estimation is done for each causally-connected group. OM hits 
which do not obey model of muon propagation and direct Cherenkov light
emission are excluded in a gradually tightening set of cuts on hit residuals.
The set of hits selected with this procedure has a noise contamination at
the level of 1-2\% depending on the elevation of the muon track. Selected set
of pulses is used for the track fit under the assumption of direct Cherenkov
emission from the muon track. The resulting median resolution of the procedure
as measured in the up-going neutrino Monte Carlo sample is at the level
of 1 degree.

Figure \ref{muons_33days} (left panel) shows the reconstructed angular
distribution in data compared to the
prediction of Monte Carlo simulation of the detector response to atmospheric
muons. Good agreement in shape of the distributions is achieved although the
muon rate in data is 1.5 times larger than in MC. Muon bundles
misrecontructed as
up-going muons constitute a large background to the up-going neutrino search.

\begin{figure}[h]
\includegraphics[width=6cm,clip]{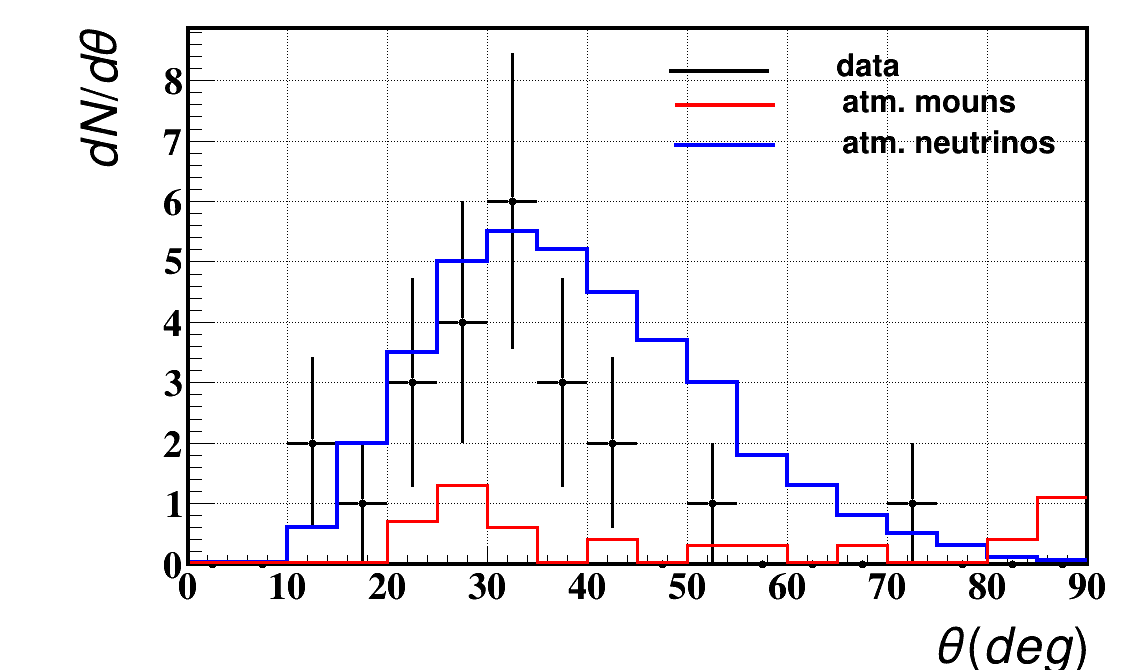}
\hfill
\includegraphics[width=6cm,height=3.5cm]{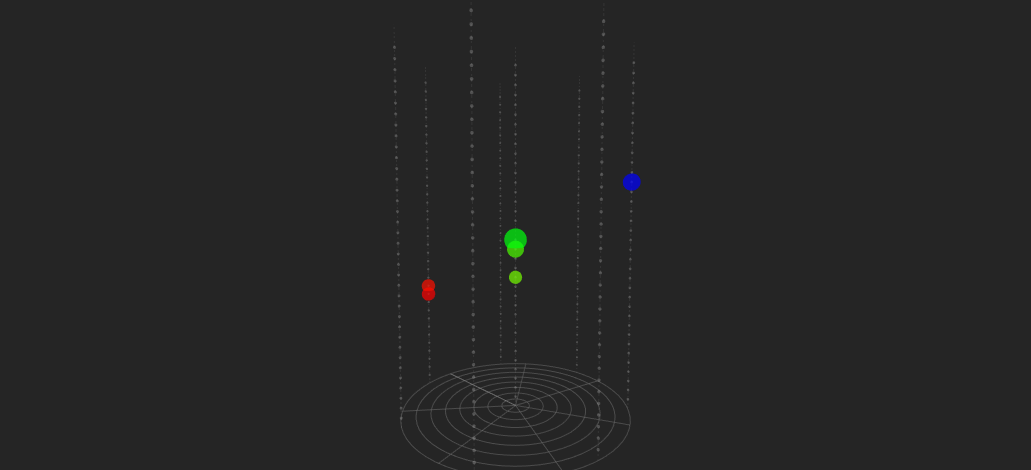}
\caption{Left: Angular distributions of selected neutrino events,
as well as expectations from atmospheric muons and neutrinos.
Right: Event view of upward moving neutrino observed in this analysis.
Each sphere represents an OM.
Colors represent the arrival times of the photons where red indicates
early and blue late times. The size of the spheres is a measure for
the recorded number of photoelectrons.}
\label{fig-2.1.2}       
\end{figure}

A procedure based on a boosted decision tree (BDT) as implemented in the TMVA
framework \cite{reftmva} was developed for the selection of neutrino events.
A set of quality variables was reconstructed for each event and used for the
BDT discriminant. The BDT was trained on events reconstructed as up-going in MC
samples of atmospheric up-going neutrinos (signal) and atmospheric muons
(background). Good signal/backgorund discrimination was achieved for a
BDT value of 0.2 (Figure 2, right panel). The BDT value was calculated
for the data events and events with a BDT value > 0.2 were selected.
In total 23 neutrino
candidate events were found (Figure \ref{fig-2.1.2}, left panel) while 42
events are expected from up-going neutrino MC. The number of expected
background
events is about 6. In Figure \ref{fig-2.1.2} (right) event view of one upward
moving neutrino observed in the present analysis is shown.

\subsection{Cascade detection by GVD}
\label{sec-2.2}
IceCube discovered a diffuse flux of high-energy
astrophysical neutrinos in 2013 \cite{IC1}.
The data sample of their high-energy starting event analysis
(HESE, 7.5 year sample) comprises 103 events, 77 of which are
identified as cascades and 26 as track events \cite{IC2}.
These results demonstrate the importance of the cascade
mode of neutrino detection with neutrino
telescopes. The Baikal Collaboration has a long-term experience
to search for a diffuse neutrino flux
with the NT200 array using the cascade mode \cite{NT1,NT2}. Baikal-GVD 
has the potential to record
astrophysical neutrinos with flux values measured by IceCube
\cite{IC3} even
at early phases of construction.
A search for high-energy
neutrinos with Baikal-GVD is based on the selection of
cascade events generated by neutrino
interactions in the sensitive volume of the array \cite{CAS1}.
Here we discuss the first preliminary results
obtained by the analysis of data
accumulated with Baikal-GVD in 2015-2016.

To search for high-energy neutrino flux of astrophysical
origin, the data collected from 24
October till 17 December 2015 have been used. A data sample
of triggered $4.4 \times 10^8$ events has been
accumulated, which corresponds
to 41.64 live days. Causality cuts and the
requirement of N $ \ge $ 3 hit OMs leave about
$1.8 \times 10^7$ events for the following analysis.
After applying an
iterative procedure of cascade vertex reconstruction
for hits with charge higher 1.5 ph.el.,
followed by the rejection of hits contradicting
the cascade hypothesis on each iteration stage,
316,229 events survived.
After cascade energy reconstruction and event
quality cuts, 12,931 cascade-like events
survive.
A total of 1192 events from final sample were
reconstructed with energies above 100 TeV. The
multiplicity distribution of of hit OMs for these events is shown in 
Figure 4 (left).
Also shown are the expected event distributions from
an astrophysical flux with an
$E^{-2.46}$ spectrum and the IceCube normalization, as well as
the expected distributions from atmospheric muons and
atmospheric neutrinos. The statistics of the generated atmospheric
muon sample correspondes to 72 live days data taking.

\begin{figure}[h]
\includegraphics[width=6cm,height=4.5cm]{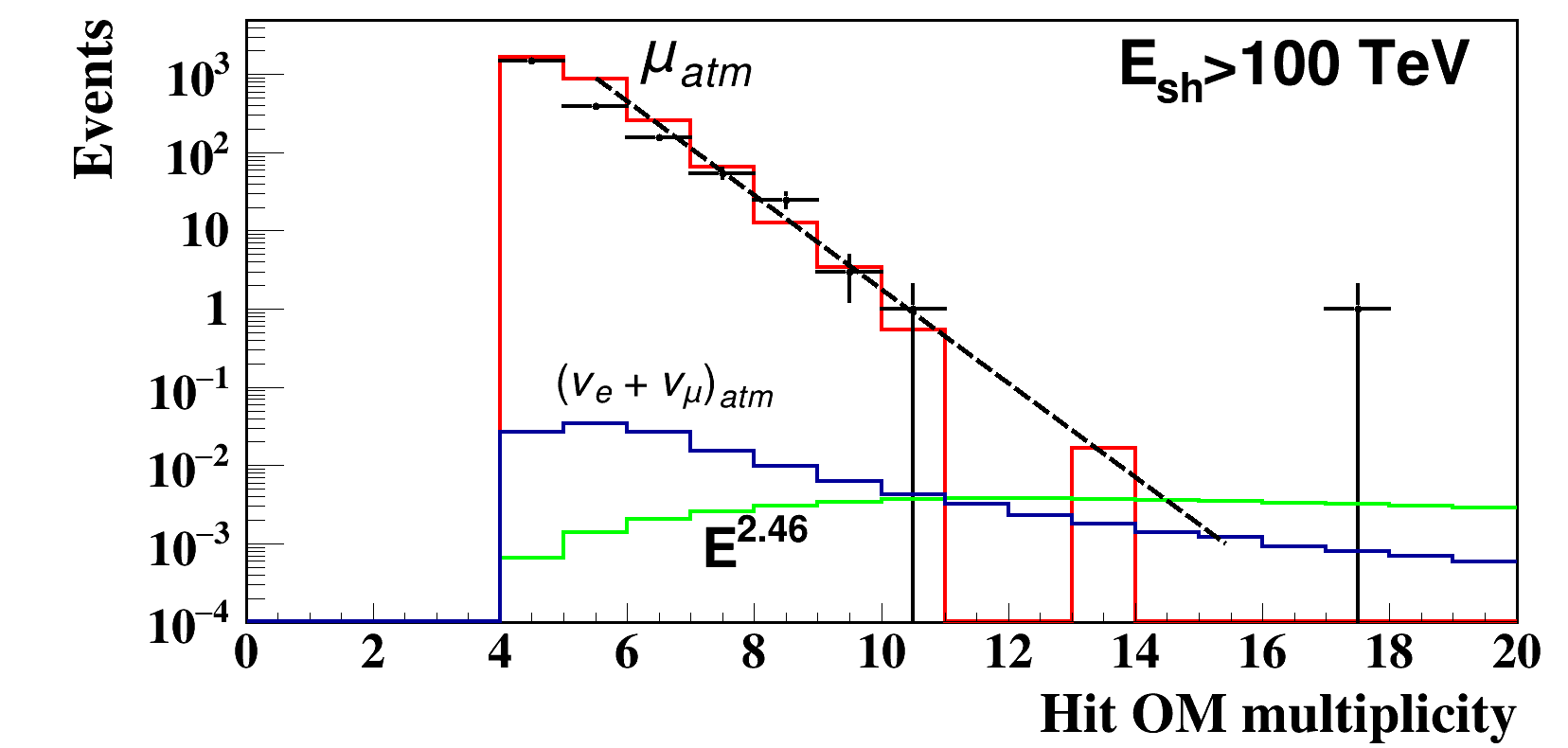}
\hfill
\includegraphics[width=9cm,height=5cm]{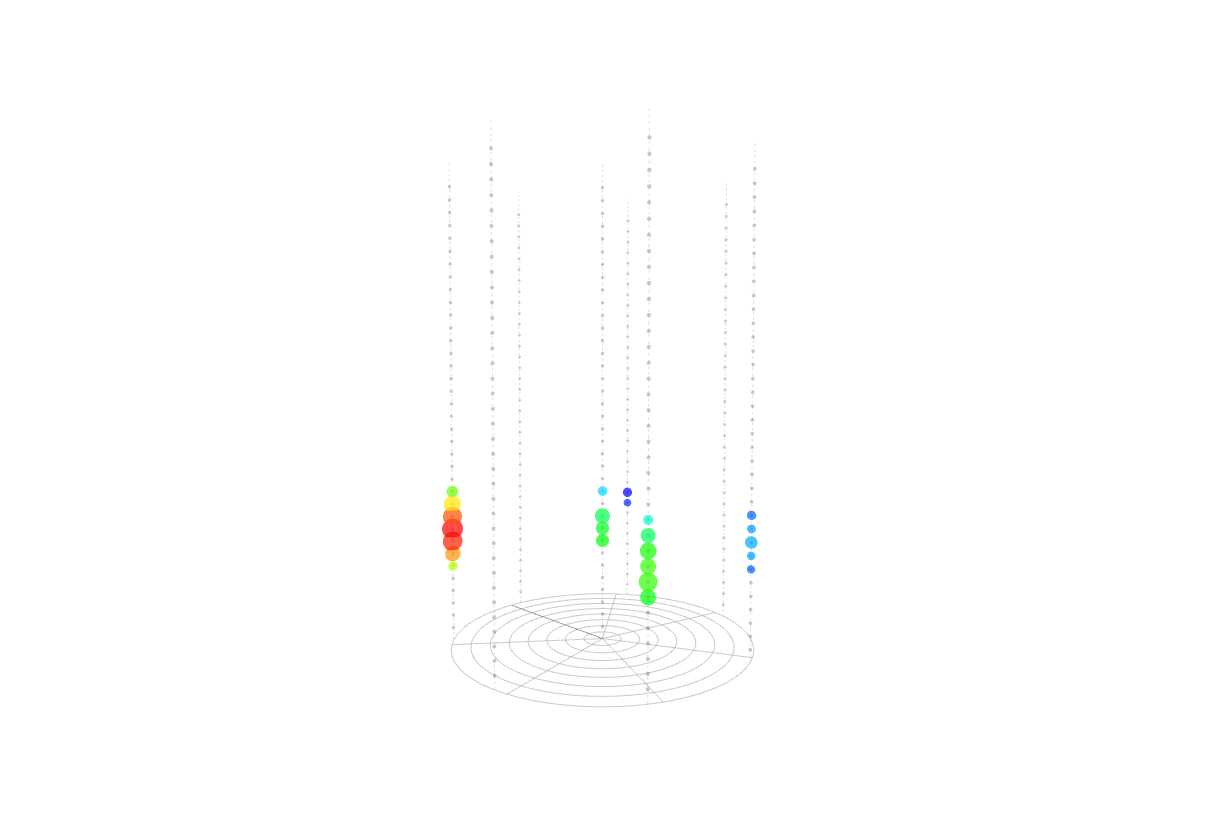}
\caption{Left: Multiplicity distribution of hit OMs for
experimental events
with reconstructed energy $E_{rec}$ > 100 TeV (dots). Also
shown are the distributions of events expected from
astrophysical neutrinos with an $E^{-2.46}$ spectrum and background
events from atmospheric muons and neutrinos.
Right: The event observed in October 2015.}
\label{fig-2.2.1}       
\end{figure}

All but one experimental events have multiplicities
less than 10 hit OMs and are
consistent with the expected number of background events
from atmospheric muons. One event with 17 hit OMs
was reconstructed as downward moving cascade.
For a more precise reconstruction of cascade
parameters, this event was reanalysed including
hits with charges lower 1.5 ph.el.. 24
hits are consistent with a cascade hypothesis
and the following cascade parameters:
cascade energy E = 107 TeV, zenith angle
$\theta = 56.6^{\circ} $
and azimuthal angle $\phi = 130.5^{\circ}$
\footnote{The reconstructed directional vector
$\vec{\Omega}(\theta,\phi)$ is
opposite to the direction of the cascade development axis in water
and represents the coordinates of a potential neutrino
source on the celestial sphere in the array coordinate system.}, distance
from the array axis $\rho = 67.7$ m. 
The event is shown in
Figure 4 (right panel).

The search for cascades from astrophysical neutrinos
has been continued with data collected
between April 2016 and January 2017, which corresponds to
an effective livetime of 182 days. A data sample
of $3.3 \times 10^8$ events was
selected after applying causality cuts and the
requirement of N $ \ge $ 3 hit OMs with hit charges $\ge$1.5 ph.el.
on $\ge $ 3 strings. 

At the next stage of the analysis the cascade reconstruction procedure
and a set of quality cuts have been applied to data.
In Table 1 the number of surviving events and the efficiency of 
applied cuts are shown. Here $\chi^2_t$ - value of the minimizing
function after cascade vertex reconstruction, $L_A$ - log 
likelihood after energy reconstruction, $\eta$ - variable which
depends on probabilities of hit OMs to be hit and non-hit OMs
not to be hit.
Positive values of $\eta$ are expected for cascades. 
Hit multiplicity distributions of events
after cuts from Table 1 are shown in Figure 5 (left). In the right panel
of Figure 5 the hit multiplicity of events with $E_{sh} > $ 10 TeV and
expected distribution of background events from atmospheric muons are shown. 
\begin{table}[h]
\centering
\caption{Efficiency of applied cuts}
\label{tab-1}       
\begin{tabular}{ccc}
\hline
Cuts & Events & Rejection factor  \\\hline
\multicolumn{3}{c}{After cascade vertex reconstruction}\\ \hline
N$_{hit} \ge$ 10 & 577495 & 1 \\\hline
$\chi^2_t <$ 4 & 2405 & 1/240 \\\hline
\multicolumn{3}{c}{After cascade energy and direction reconstruction}\\ \hline
$L_A <$ 20 & 374 & 1/6.4 \\\hline
$\eta >$ 0 & 159 & 1/2.4 \\\hline
$E_{sh} >$ 10 TeV& 57 & 1/2.8 \\\hline
$E_{sh} >$ 100 TeV& 5 & 1/11.4 \\\hline
Total rejection factor: &  & 1/115499 \\\hline 
\end{tabular}
\end{table}
Finally, 57 events with reconstructed energies $E_{sh} > $ 10 TeV and
5 events with $E_{sh} > $ 100 TeV have been selected. 
Four of five events with energies higher than 100 TeV
have hit multiplicities consistent with the expected
distribution of background events from atmospheric muons.
\begin{figure}[h]
\includegraphics[width=6cm,height=5cm]{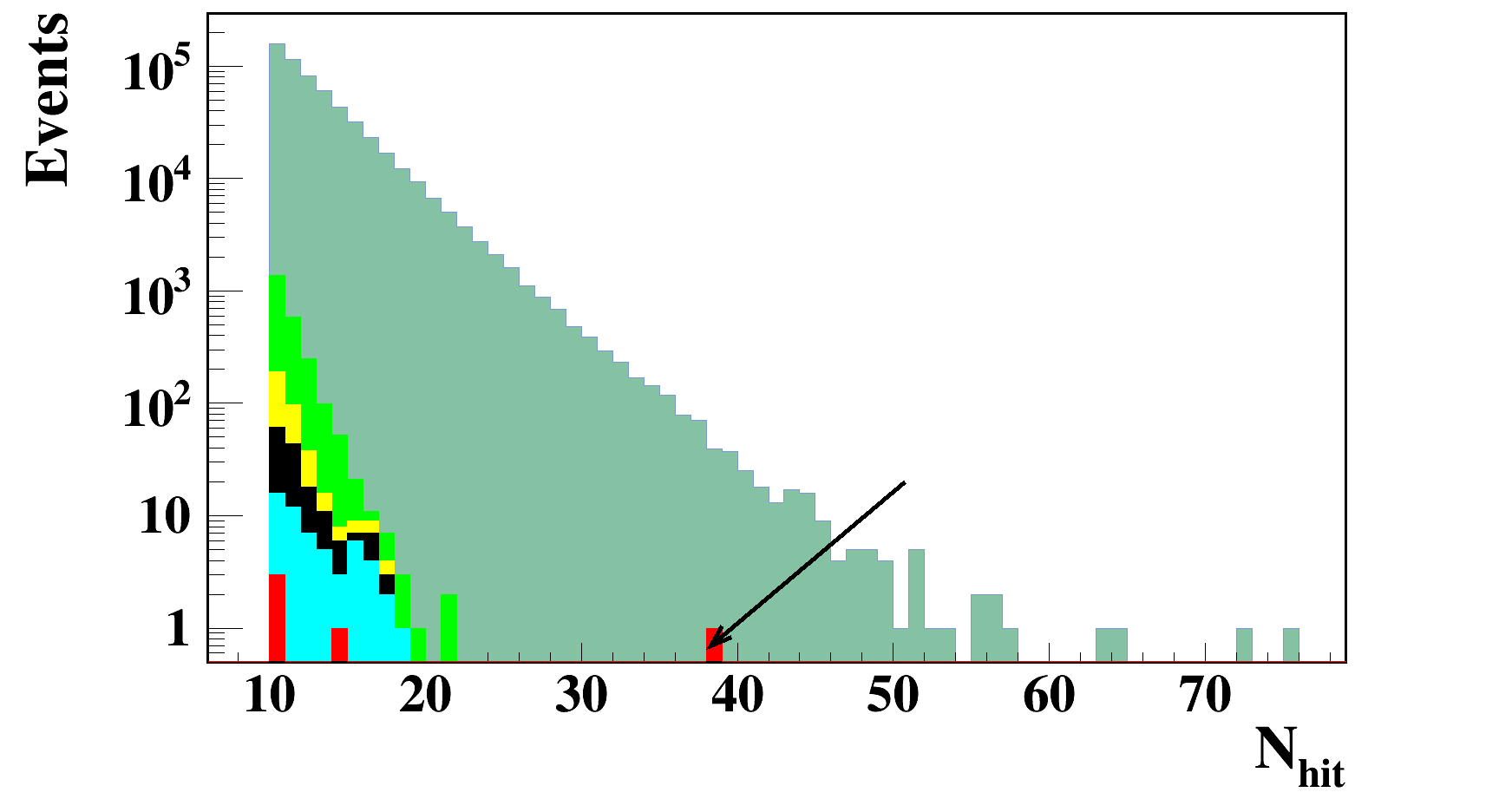}
\hfill
\includegraphics[width=6cm,height=5cm]{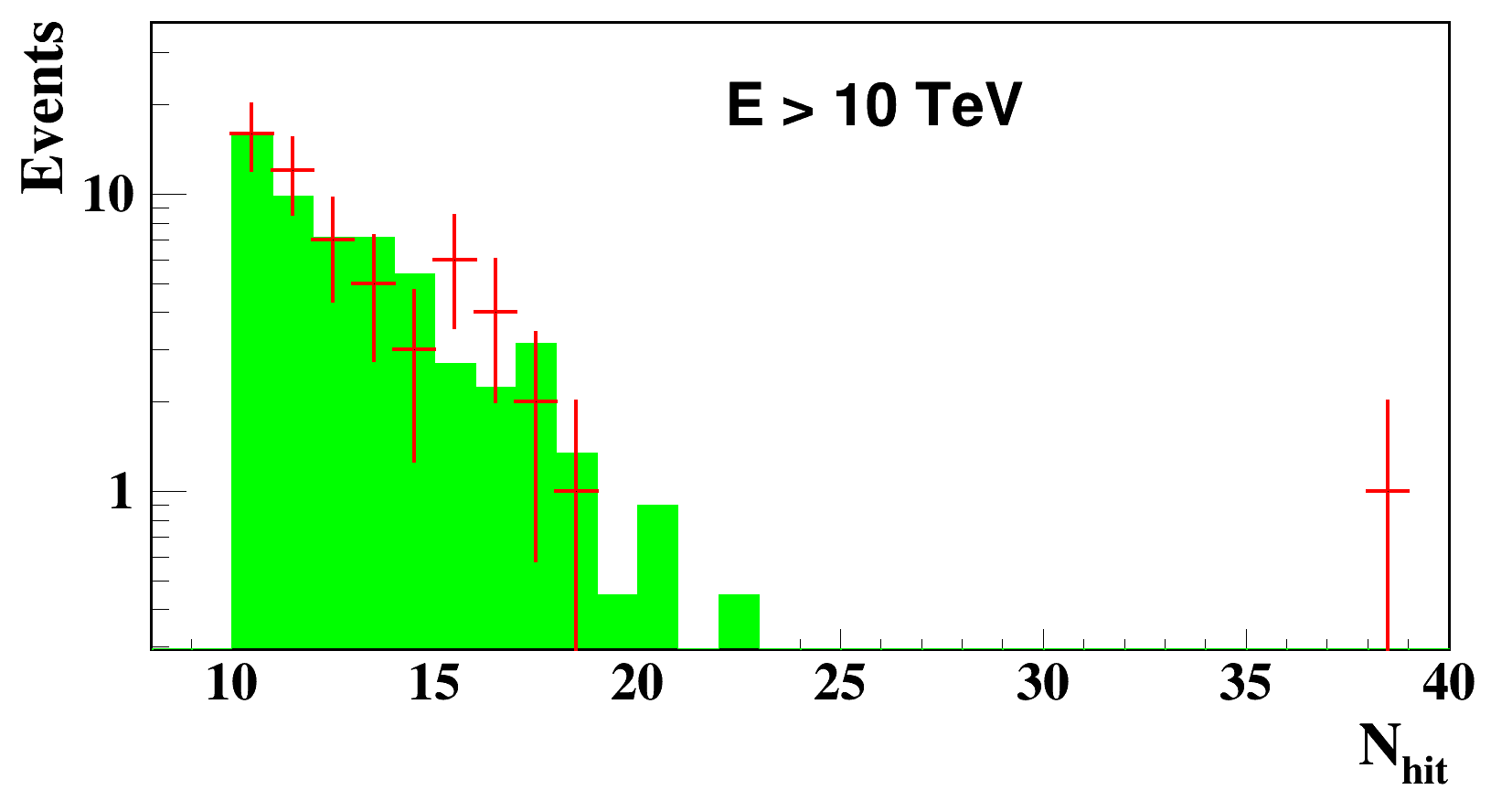}
\caption{Left: Multiplicity distributions of hit OMs after cuts
explained in Table 1. Right: The same 
for data (points) and atmospheric muons (histogram) with
reconstructed energies $E_{sh} >$ 10 TeV.
}
\label{fig-2.2.2}       
\end{figure}
\begin{figure}[h]
\includegraphics[width=6cm,height=5cm]{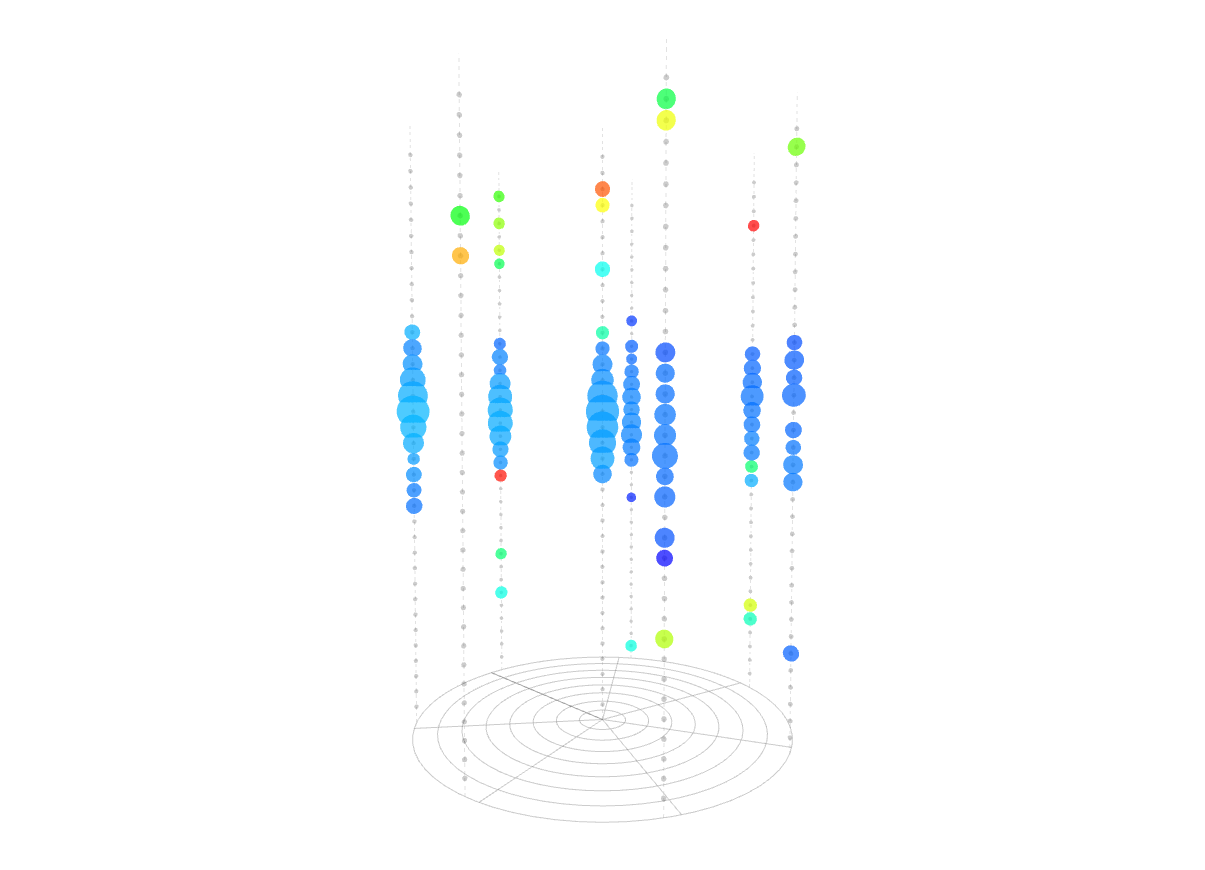}
\hfill
\includegraphics[width=6cm,height=5cm]{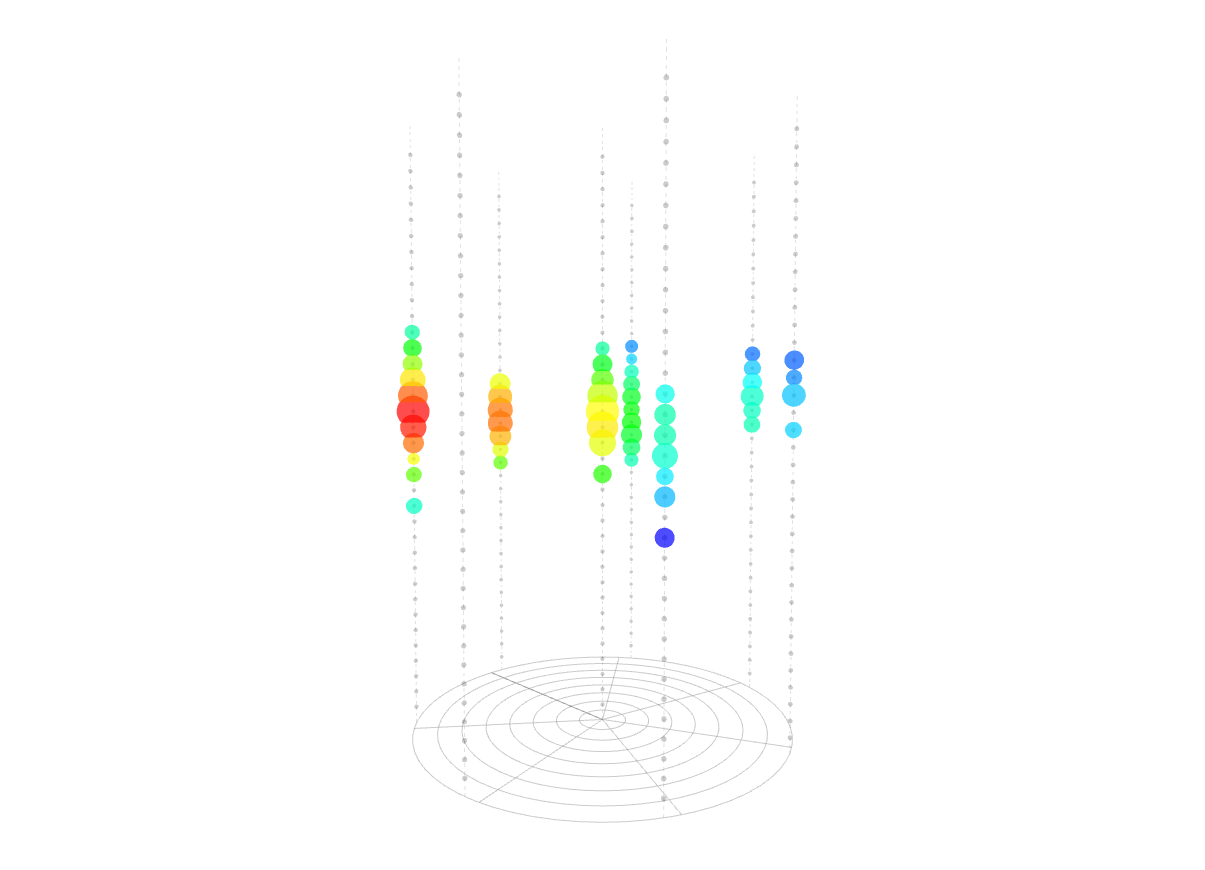}
\caption{The event observed in April 2016: left -
all hit OMs, right - hits which survive all cuts.}
\label{fig-2.2.3}       
\end{figure}
One event with 38 hit OMs
was reconstructed as downward moving contained cascade.
For more precise reconstruction of cascade
parameters, this event was reanalysed including
hits with charges lower 1.5 ph.el.. 53
hits are consistent with a cascade hypothesis
and the following parameters:
energy E = 154.9 TeV, zenith angle
$\theta = 57.3^{\circ} $ and
azimuthal angle $\phi = 249.4^{\circ}$, distance
from the array axis $\rho = 44.7$ m. The event is shown in
Figure 6. In the left and right panels hit OMs before and after
noise hit rejection are shown, respectively.

The two clear high-energy cascade events have been selected
from data recorded during 2015-2016. Coordinates of the potential
sources of these events are the following:
right ascension (RA) 229.5$^{\circ}$
and declination (Dec) 5.6$^{\circ}$
for the first (2015) event, and right ascension (RA) 173.4$^{\circ}$
and declination (Dec) 13.9$^{\circ}$
for the second (2016) event in equatorial coordinates.

\subsection{Search for high-energy neutrinos associated with GW170817}
\label{sec-2.3}
On August 17, 2017, a gravitational wave signal, GW170817, from a
binary neutron star merger has been recorded
by the Advanced LIGO and Advanced Virgo
observatories \cite{GW}.
A short GRB (GRB170817A), associated with GW170817, was detected
by Fermi-GBM and INTEGRAL. Optical observations allowed the
precise localization of the merger in the galaxy NGC 4993 at
a distance of $\sim$40 Mpc. High-energy neutrino signals
associated with the merger were searched for by the ANTARES
and IceCube neutrino telescopes in muon and cascade modes
and the Pierre Auger Observatory \cite{GWNU_1} and
Super-Kamiokande \cite{GWNU_2}. Two different time windows
were used for the searches. First, a $\pm$500 s time
window around the merger was used to search for neutrinos associated
with prompt and extended gamma-ray emission \cite{Baret,Kimura}.
Second, a 14-day time window following the GW detection, to cover
predictions of longer-lived emission processes \cite{Gao,Fang}.
No significant neutrino signal
was observed by the neutrino telescopes.

\begin{figure}[ht]
\includegraphics[width=7cm,height=6.5cm]{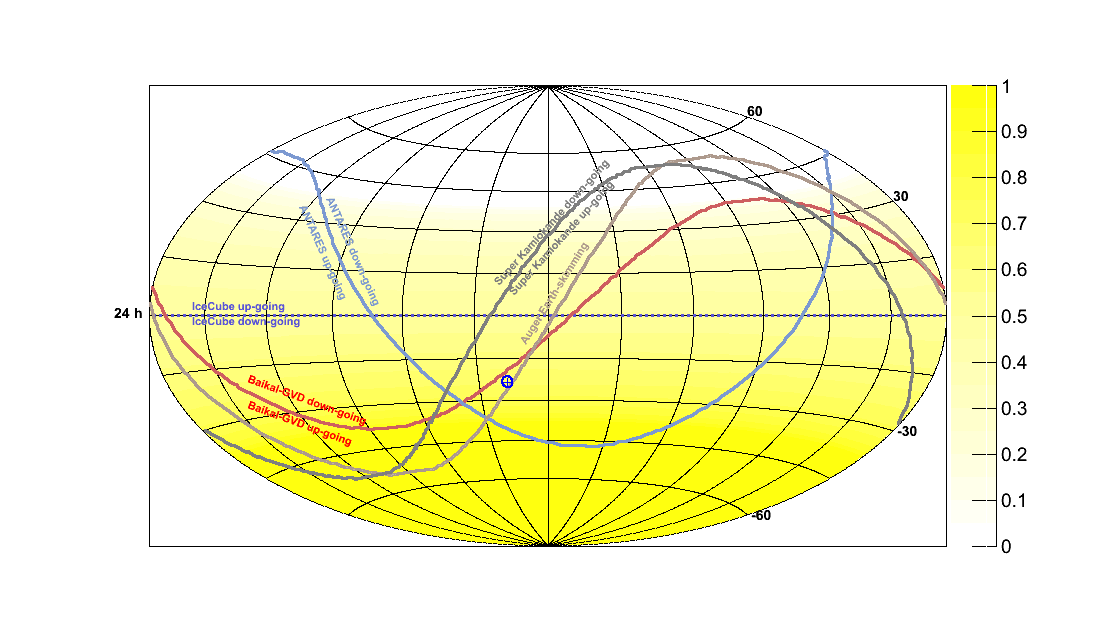}
\hfill
\includegraphics[width=6cm,height=5.8cm]{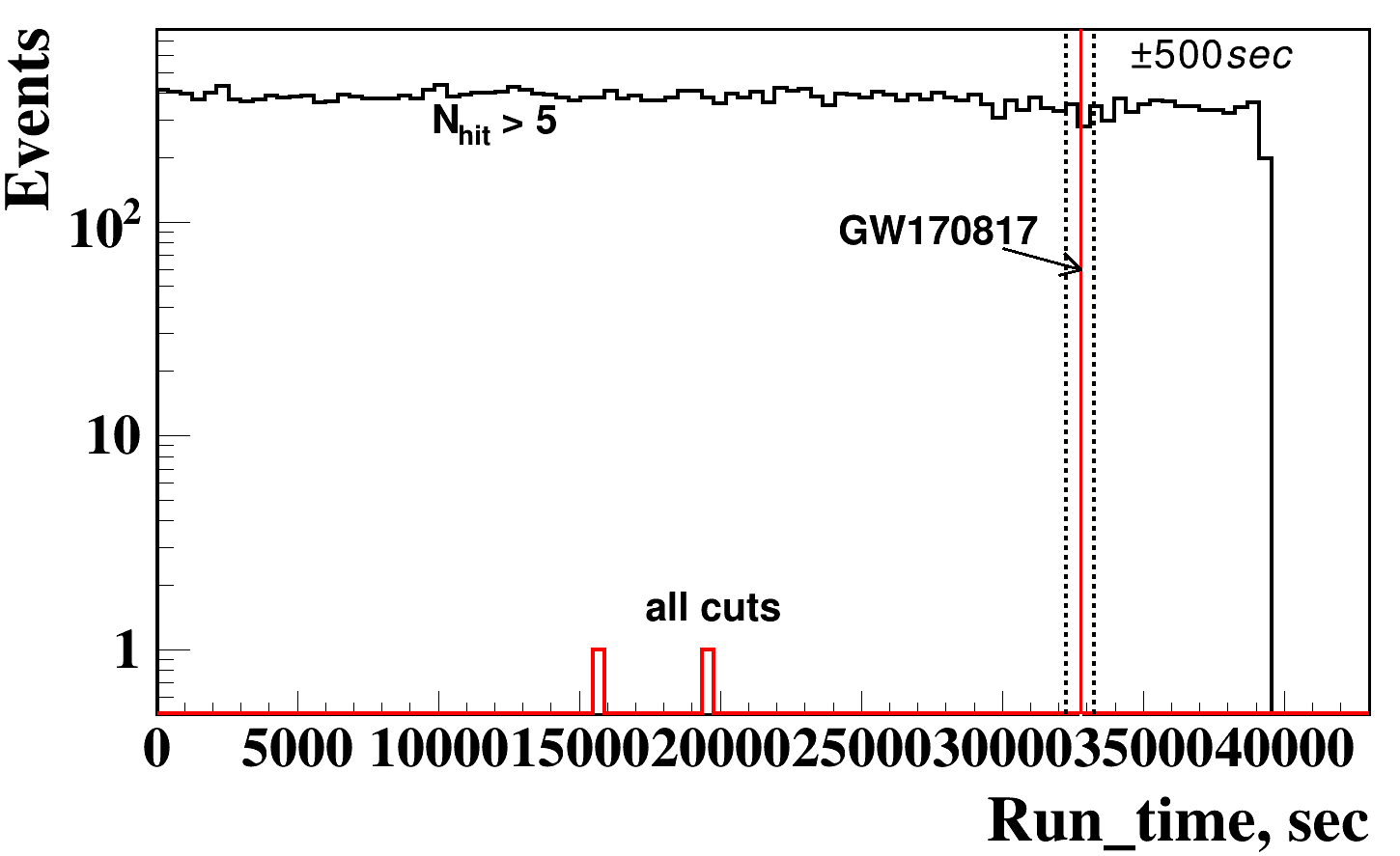}
\caption{Left: Localizations of NGC 4993 and horizons separating down-going
and up-going neutrino directions
for IceCube, ANTARES, SuperKamiokande and Baikal-GVD at the
time of the GW event in equatorial coordinates.
The zenith angle of the source at the
detection time of the merger was 73.8$^{\circ}$ for ANTARES,
66.6$^{\circ}$ for IceCube,
108$^{\circ}$ for SK and 93.3$^{\circ}$ for Baikal-GVD.
Right: Temporal distribution of events during the data taking run contaning
the $\pm$500 s time window around the GW event. The black histogram represents
events with hit OMs N$_{hit}>$5, and the red histogram represents events
surviving all selection cuts used for the neutrino search within $\pm$500 s
time window around the GW event.   
}
\label{fig-2.3.1}       
\end{figure}

Here we discuss preliminary results of a search for high-energy
neutrinos in coincidence with GW170817/GRB170817A using the cascade
mode of neutrino detection. Two GVD-clusters have been operated during 2017.
The zenith angle of NGC 4993 at the detection time
of GW170817 was 93.3$^{\circ}$ for Baikal-GVD
(see, Figure \ref{fig-2.3.1}, left panel). Since background
events from atmospheric muons and neutrinos can be drastically
suppressed by requiring time and space coincidence with the GW signal,
relatively weak cuts can be used for neutrino selection. For the search for
neutrino events within $\pm$500 s window around the GW event, 731 events were
selected, which comprise >5 hit OMs at >2 hit strings. After applying
cascade reconstruction procedures and dedicated quality cuts, two events
were selected. Finally, requiring directional coincidence with NGC 4993
$\psi < 20^{\circ}$ no neutrino
candidates survive. The median angular error is
4.5$^{\circ}$ with this set of relaxed cuts and
the expected number of atmospheric background
events is about $5 \times 10^{-2}$ during the coincident time window.
Shown in
Figure \ref{fig-2.3.1} (right panel) are temporal distributions of events
fulfilling the additional selection requirement (black histogram) as well as
events surviving all cuts (red histogram) during the 39347 s long data
taking run, which containes the $\pm$500 s time window around GW170817.

\begin{figure}[h]
\centering
\sidecaption
\includegraphics[width=8cm,height=7.5cm]{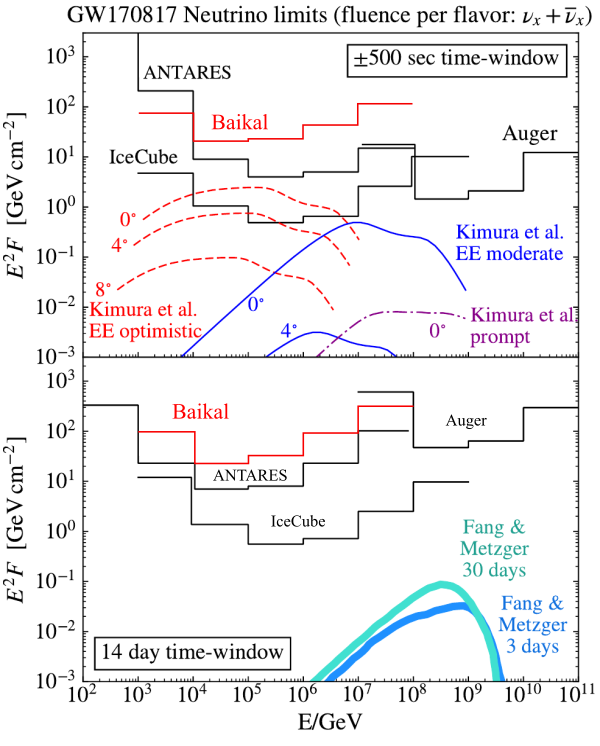}
\caption{Upper limits (at 90 \% confidence level) on the neutrino
spectral fluence from GW170817 during a $\pm500$ s window centered
on the GW trigger time (top panel), and a 14-day window following
the GW trigger (bottom panel). For each experiment, limits are
calculated separately for each energy decade, assuming a spectral
fluence $F(E) = F_{up} \times [E/GeV]^{−2}$ in that decade only. Also
shown are predictions by neutrino emission models (see \cite{GWNU_1}
for details). 
}
\label{fig-2.3.3}       
\end{figure}

The search over 14 days used a more stringent cut on the number of
hit OMs - N$_{hit} >$7. No events
spatially coincident with GRB170817A were found. Given the non-detection
of neutrino events associated with GW170817, upper limits on
the neutrino fluence have been derived (see Figure \ref{fig-2.3.3}). 
 
\section{Conclusion}

The ultimate goal of the Baikal-GVD project
is the construction of a km3-scale neutrino telescope with
implementation of about ten thousand light sensors.
The array construction was started
by deployment of reduced-size demonstration cluster named "Dubna"
in 2015, which comprises 192 optical modules.
The first cluster in it's baseline configuration was deployed
in 2016 and the second one in 2017. After deployment of the third
GVD-cluster in April 2018 Baikal-GVD comprises the total of 864 OMs
arranged at 24 strings and becomes, at present, the largest
underwater neutrino telescope. The modular structure of Baikal-GVD
design allows  studies of neutrinos of
different origin with early stages of construction.
Analysis of data collected in 2015-2017 allows for extraction
of a sample of upward through-going muons as clear neutrino candidates
and the identification
of the first two promising high-energy cascade events -
candidates for events from astrophysical neutrinos. The search for
neutrinos assosiated with GW170817 with Baikal-GVD
allows to derive upper limits on
the neutrino spectral fluence from this source. 
The commissioning of the
first stage of the Baikal neutrino telescope GVD-1 with an
effective volume 0.4 km$^3$ is envisaged for 2020-2021. 

\begin{acknowledgement}
{\it This work was supported by the Russian Foundation for Basic
Research (Grants 16-29-13032, 17-02-01237).}
\end{acknowledgement}

%
%

\end{document}